\documentclass{article}
\usepackage[T1]{fontenc} 
\usepackage[utf8]{inputenc} 
\usepackage{ismir,amsmath,cite,url}
\usepackage{graphicx}
\usepackage{color}
\usepackage{microtype}
\usepackage{units}
\usepackage{paralist}
\usepackage{svg}
\usepackage{pdfpages}
\usepackage{subcaption}


\usepackage{lineno}
\graphicspath{ {./figures/} }
\title{Improving Music Performance Assessment with Contrastive Learning}

\twoauthors
    {Pavan Seshadri} {Center for Music Technology \\ Georgia Institute of Technology \\ {\tt{pseshadri9@gatech.edu}}}
    {Alexander Lerch} {Center for Music Technology \\ Georgia Institute of Technology\\ {\tt{alexander.lerch@gatech.edu}}}

\def\authorname{P.\ Seshadri and A.\ Lerch}

\begin{document}

\maketitle
\begin{abstract}
Several automatic approaches for objective music performance assessment (MPA) have been proposed in the past, however, existing systems are not yet capable of reliably predicting ratings with the same accuracy as professional judges. This study investigates contrastive learning as a potential method to improve existing MPA systems. Contrastive learning is a widely used technique in representation learning to learn a structured latent space capable of separately clustering multiple classes. It has been shown to produce state of the art results for image-based classification problems. We introduce a weighted contrastive loss suitable for regression tasks applied to a convolutional neural network and show that contrastive loss results in performance gains in regression tasks for MPA. 
Our results show that contrastive-based methods are able to match and exceed SoTA performance for MPA regression tasks by creating better class clusters within the latent space of the neural networks.
\end{abstract}
\section{Introduction}\label{sec:introduction}

%
Within the context of western classical music, musical performances are a sonic interpretation of a written musical score. Performers are 
tasked with interpreting the score 
and translating it to an acoustic rendition. In doing so, they craft a unique performance by
controlling and varying performance parameters such as tempo 
and timing, dynamics, intonation, and tone quality \cite{clarke_understanding_2002}. These performance parameters and their variation impact the way in which listeners perceive the music, and let them distinguish between performances of the same musical score \cite{lerch_interdisciplinary_2020}. 

For music performers, the journey to competence and mastery often spans years of practice and tailored instruction. As music performance is an inherently complex and subjective task, proper feedback on performances is imperative to growth as a performer, and such, regular feedback by professional musicians is necessary. Music teachers are expected to evaluate students on various criteria such as musicality or tone quality, although rating these criteria is highly subjective, complicating the task of consistent and objective Music Performance Assessment (MPA) \cite{wesolowski_examining_2016, thompson_evaluating_2003}. These challenges, however, do not reduce the need of assessing music performances, e.g., in institutions of musical education. Thus, any effort towards either formalizing human assessments or the creation of objective, reproducible, and unbiased systems for automatic assessment contributes to overcoming the above-mentioned challenges. 

A system for automatic MPA can be used for software-based music tutoring applications to allow for easier accessibility of music education and individualized instruction. Past this, such objective assessment systems might also serve as tools for the evaluation of performance generation systems. 



The approaches in automatic MPA follow the same general historical patterns as other audio analysis systems. Older systems extract hand-crafted features from recorded performances and then use a data-driven approach such as a regression model to map the features to a grade or assessment rating that reflects human ratings \cite{wu_towards_2016}. Deep learning methods have since been found to outperform feature extraction-based methods \cite{pati_assessment_2018}; however, modern systems still fall short of the reliability required for a ready-to-use system \cite{huang_score-informed_2020}.

Representation learning aims to accurately encode relevant and useful characteristics into a compressed representation. Representation learning methods such as VGGish have been shown to encode powerful audio features into a compressed representation which~---when used as input to classification systems---~can produce state of the art performance \cite{hershey_cnn_2017}. 
Contrastive learning is an emerging representation learning method which uses a distance-based loss between pairs of encoded training points in order to create meaningful class separation within the latent space of a neural network \cite{hadsell_dimensionality_2006}. 

This study aims to investigate the use of contrastive learning to improve the performance of MPA systems. We investigate the use of contrastive-based learning methods in a regression task, where a deep neural network taking an input audio recording of a musical performance is tasked with estimating a numerical rating consistent with that of a professional judge. Our hypothesis is that learning a structured latent space will improve the ability of the regression components of MPA models in predicting an assessment rating.
We investigate two methods of incorporating contrastive learning into a standard CNN-based architecture to learn a structured latent space,
\begin{inparaenum}[(i)]
    \item   a two-step training method introduced by Khosla et al.\ \cite{khosla_supervised_2020}, and 
    \item   a joint loss method combining the contrastive loss term with a mean squared error loss term, a standard loss for training regression systems. 
\end{inparaenum}
As contrastive loss within a supervised context is generally designed for classification tasks \cite{khosla_supervised_2020} as opposed to regression tasks, we introduce a weighted contrastive loss suitable for regression tasks.


The remainder of this paper is structured as follows. First, we give an overview of music performance assessment and previous work on contrastive loss. Then, we introduce the proposed method in Sect.~\ref{sec:method}. Section~\ref{sec:experiment} introduces our experimental setup. In the following Sect.~\ref{sec:result}, we evaluate the performance of the contrastive-based methods against a baseline architecture to predict a regression rating and perform analysis on the latent space of the baseline and contrastive-based methods to determine the efficacy of this clustering on the overall performance. Overall, we find that contrastive-based learning is able to better cluster the latent representation and produce performance gains within our MPA regression task.

\section{Related Work}
MPA aims to understand and 
model the parameters of a musical performance and investigate their impact on a human listener \cite{lerch_software-based_2009}. MPA systems thus have the goal of assessing musical performances based on audio recordings without the input of expert judges.   
Early research on musical performances centered around analyzing symbolic data extracted from MIDI devices \cite{palmer_mapping_1989, repp_patterns_1996}, whereas recent research has increasingly focused on 
analyzing raw audio \cite{dixon_pinpointing_2002,lerch_software-based_2009}. In human performance assessment, music 
instructors must discern the individual subjective qualities and criteria and their importance. Similarly, automatic performance assessment systems extract features representing the audio file and then use a data-driven model to estimate the assessment rating. The features are either hand-crafted for the task \cite{nakano_automatic_2006, knight_potential_2011, dittmar_music_2012,wu_towards_2016, gururani_analysis_2018}, or learned from the training data \cite{han_hierarchical_2014, pati_assessment_2018, wu_learned_2018}. 
Systems with handcrafted features often use traditional machine learning approaches \cite{vidwans_objective_2017} while feature learning is usually done within a more complex neural model with low-level input representations such as spectrograms \cite{7953122}.

Some studies specifically aim to automatically produce a numerical rating on a predefined scale from audio representations of a musical performance, which involves implicitly learning the aspects of performances that correlate to certain rating criteria \cite{wu_learned_2018, pati_assessment_2018, huang_score-informed_2020}. However, since numerical scores do not inherently include specific performance feedback, understanding the impacting factors can be challenging. 
The methods based on deep neural networks, while generally yielding superior performance, usually lack interpretability. Learning a structured latent space is a first step towards having a more easily understandable representation.   Representation learning is an emerging method for performance assessment. For example, Huang et al.\ proposed a joint-embedding network which learns a shared latent space of a performance and its written score and derives a regression rating by the cosine similarity between the two embeddings \cite{huang_score-informed_2020}. Representation learning methods thus potentially provide both performance improvements, as well as better interpretability of numerical scoring models, such as the ones in this study. 

An emerging method of representation learning is the Contrastive Loss. Contrastive Loss aims to regularize the latent space so that the distances between latent vectors are meaningful. This is achieved by comparing the distances between the latent representations of pairs of training points and pushing them within a set distance in the latent space if they have similar labels, and outside this set distance if they are dissimilar. These distances are compared within the contrastive loss function of a model in order to encode the information within the latent vectors. This ideally creates class clusters within the latent space. Contrastive-based loss functions are often used to specifically learn structured latent representations of data \cite{hadsell_dimensionality_2006, khosla_supervised_2020, ferraro_enriched_2021}, which then can be adapted for downstream tasks by training classifiers on these produced latent vectors \cite{khosla_supervised_2020, ferraro_enriched_2021}.
There has been considerable work done on the use of a supervised contrastive loss to cluster latent spaces for classification tasks. Chopra et al.\ introduce the max margin contrastive loss, and discuss its potential to discriminate classes when the exact number of classes may not be known, such as within recognition or verification tasks \cite{hadsell_dimensionality_2006}. Khosla et al.\ investigate a supervised contrastive loss to train deep neural networks for classification tasks on the ImageNet dataset, and found that it outperforms general cross entropy based methods \cite{khosla_supervised_2020}. This implies that using the contrastive loss can produce an advantageous latent space layout more suitable for the following tasks. Ferraro et al.\ investigate the use of contrastive learning for music and audio for three downstream MIR tasks, genre classification, playlist continuation, and automatic tagging and found that contrastive-based learning outperforms the baseline within all three tasks and achieves comparable performance to SoTA \cite{ferraro_enriched_2021}. Their findings suggest that contrastive learning is able to cluster similar musical recordings within the latent space of deep neural networks. To our knowledge, the use of contrastive learning has not been investigated within the context of MPA. Since it has been found to be advantageous within classification tasks across several modalities \cite{hadsell_dimensionality_2006, khosla_supervised_2020, ferraro_enriched_2021}, we study the application of contrastive learning to MPA.

\begin{figure}
\centering
\includegraphics[width=\columnwidth]{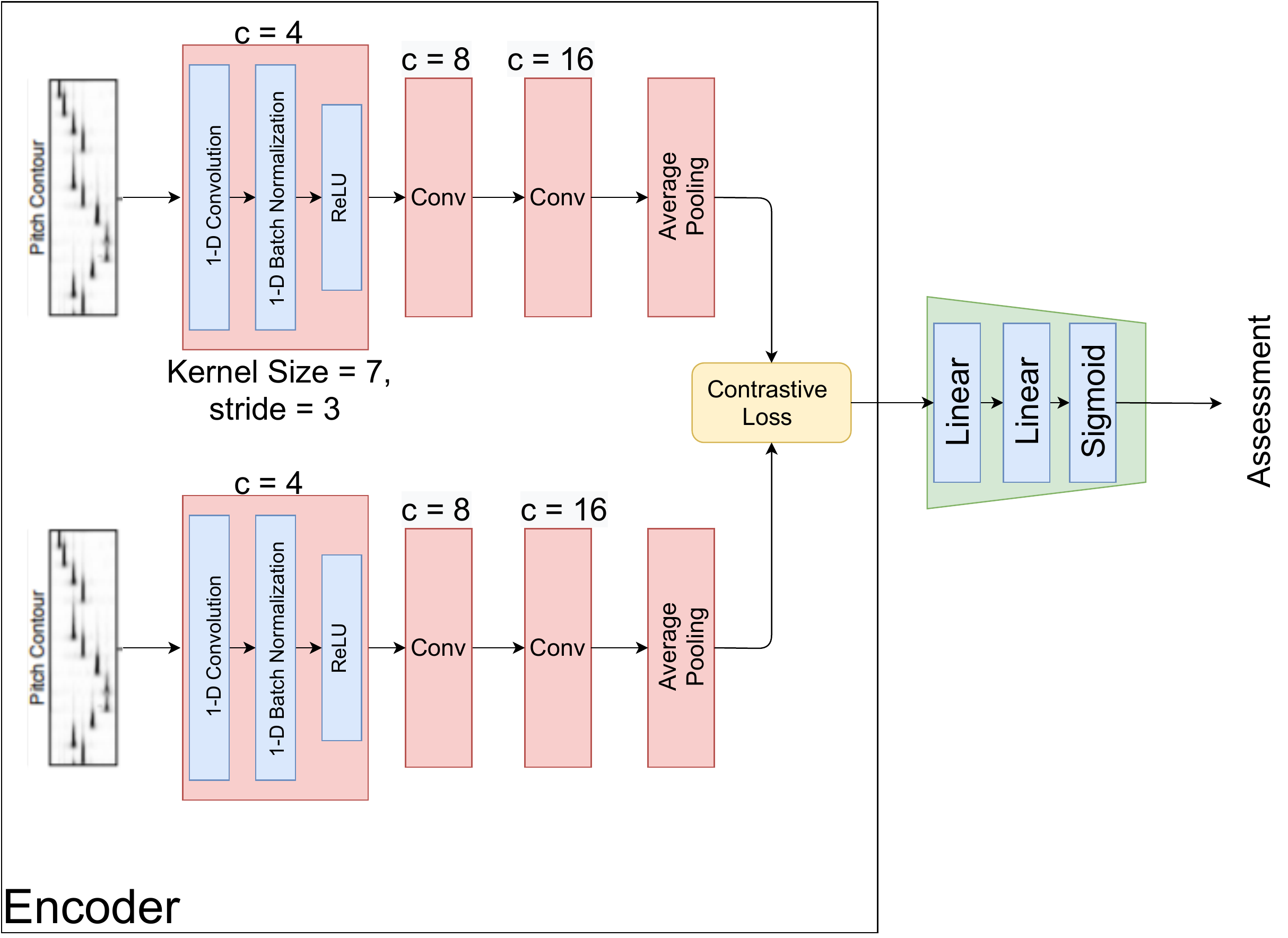}
\caption{Contrastive-Based network architecture for regression.}
\label{fig:reg}
\end{figure}

\section{Method}\label{sec:method}
We propose a weighted contrastive loss as a modification of the max margin contrastive loss introduced by Chopra et al.\ \cite{hadsell_dimensionality_2006}. The loss function is adapted to be suitable for regression tasks. We investigate incorporating the contrastive loss via two different training scenarios for a  convolutional neural network architecture. \footnote{The code is available at: https://github.com/pseshadri9/contrastive-music-performance-assessment, last accessed 8/3/2021.}

\subsection{Network architectures}
\subsubsection{Baseline}
The baseline network used is the PCConvNet architecture introduced by Pati et al.\ \cite{pati_assessment_2018}. This architecture takes pitch contours as input and uses three convolutional layers followed by an average pooling layer in order to predict assessment ratings. Each convolutional layer contains a 1-D convolution, 1-D batch normalization \cite{ioffe_batch_2015}, and ReLU non-linearity. 

\subsubsection{ContrastiveCNN}
Based on the baseline system, we design the network visualized in \figref{fig:reg}. Each branch of the network uses the same convolutional layers of the baseline with shared weights.
Two linear layers are appended to this to predict the final rating with a sigmoid activation. 

A two-step training procedure as detailed by Khosla et al.\ \cite{khosla_supervised_2020} is used to train this model. First, the encoder is trained using contrastive loss over the output latent vectors. After this, the encoder weights are frozen, and the linear layers are trained to regress this space using a mean squared error loss. Each datapoint within a training pair is fed through one encoder channel. 

\subsubsection{ContrastiveCNN-JL}
The architecture of this network is equivalent to the ContrastiveCNN, but differs in training procedure.
Rather than the two-step training procedure outlined above, the loss $L$ is the addition of the contrastive loss $L_\mathrm{C}$ over the latent vectors and the mean squared error loss $L_\mathrm{MSE}$
\begin{equation}
    L = L_\mathrm{MSE} + L_\mathrm{C} .
\label{eq:loss}
\end{equation}

\subsubsection{Input representation}
The input for each model is a pitch contour of each individual audition. Each pitch contour is an $N \times 1$ vector representing the fundamental frequency of each chunk of a performance of sequence length $N$ chunks. Pitch Contour representations were extracted using the pYIN algorithm \cite{mauch_pyin_2014} at a sample rate of \unit[44.1]{kHz} with a block size and hop size of 1024 and 256
samples, respectively. The extracted fundamental frequencies are converted to MIDI pitch values and normalized to a range of [0,1] by dividing by 127, the maximum MIDI value.

\subsection{Weighted Contrastive Loss}
Contrastive loss is generally used for classification tasks in order to create distinct class boundaries within the latent space \cite{khosla_supervised_2020, jaiswal_survey_2021, ferraro_enriched_2021, hadsell_dimensionality_2006}.
The standard max margin contrastive loss\cite{hadsell_dimensionality_2006} is defined as: 
\begin{equation}\label{contrastiveloss}
L_\mathrm{C} = \frac{1}{2}YD^{2} + \frac{1}{2}(1-Y)\max(0, (m - D))^{2} ,
\end{equation}
where $Y=1$ if the two datapoints in the pair have the same ground truth label, and $Y = 0$ if they do not. $D$ is the Euclidean distance between the two latent vectors, and $m$ is a set margin distance for which similarly labeled points should be clustered within. This results in points from the same class clustered together, while differently labeled points will be pushed past this pre-defined distance margin. 

Since this loss is not suitable for regression tasks like ours, we propose a weighted contrastive loss term. For this new loss, we first split our continuous regression range  [0, 1] into
$C$ evenly spaced rating bins. For $C=5$, for example, each rating bin has a range of 0.2, with exact multiples of 0.2 serving as the lower bounds for each bin $X$ (i.e., $[0, 0.2)$, $[0.2, 0.4)$,\ldots). Each datapoint is assigned its respective bin according to its ground truth rating. These bins are then assigned the class indices $[0,1,2,\ldots,C-1]$, which will be used for our weighted contrastive loss. 
Second, we propose a variable margin to represent the ordered nature of the rating bins. For example, it is expected that the rating bin spanning $[0, 0.2)$ should have a greater distance from the bin covering $[0.8, 1]$ than from the $[0.2, 0.4)$ bin, as the bins themselves express a rating distance. The variable margin can therefore be defined as 
\begin{equation}\label{margin}
m = |X_i - X_j|\cdot s ,
\end{equation}
where $X_i$ and $X_j$ represent the ground truth class indices of each datapoint within a pair $(X_i, X_j)$ and $s$ is the set margin distance. This variable margin scales the set distance proportionally to the expected distance between each rating bin.
This variable margin then replaces the fixed margin $m$ in Eq.~(\ref{contrastiveloss}).

\begin{table}
 \begin{center}
 \begin{tabular}{l|c|c}
    & \textbf{{Middle School}} & \textbf{{Symphonic Band}}\\
  \hline
  \textbf{Alto Sax}  & 696 & 641 \\
  \hline
  \textbf{Clarinet}  & 925 & 1156 \\
  \hline
  \textbf{Flute} & 989 & 1196 \\
 \end{tabular}
\end{center}
 \caption{Number of recordings per instrument.}
 \label{tab:datapoints}
\end{table}


\section{Experiments}\label{sec:experiment}
Our experiments investigate primarily the performance of the proposed contrastive-based methods for MPA. In particular, we are interested in evaluating
\begin{inparaenum}[(i)]
    \item   the raw performance in predicting ratings, 
    \item   the quality of the clustering of the latent spaces, and 
    \item   the effect of this clustering on the performance.
\end{inparaenum}
We evaluate the learned representation both quantitatively and qualitatively.

\begin{figure}
\centering
\includegraphics[width=\columnwidth]{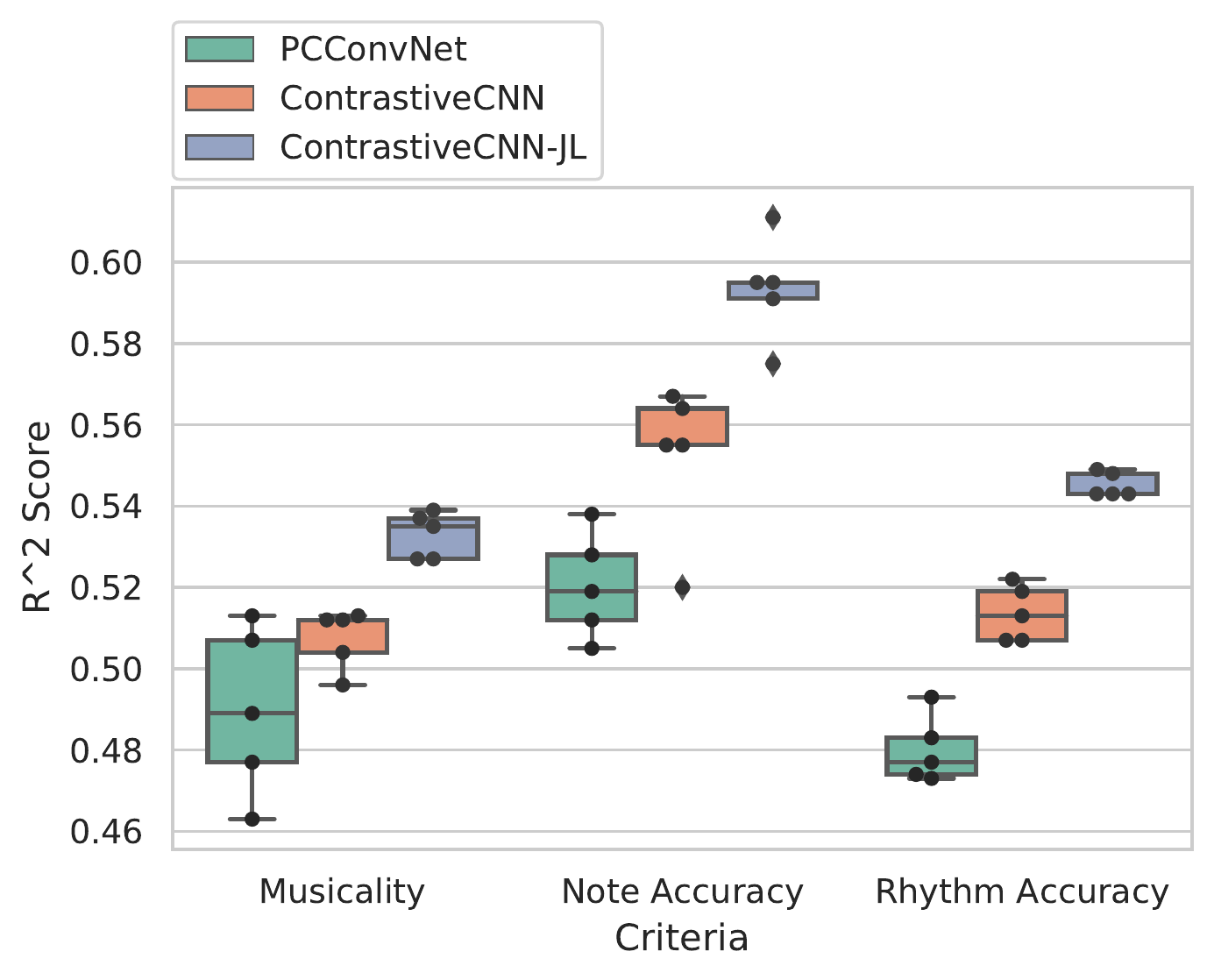}
\caption{Results over the \textit{{Middle School}} set.}
\label{fig:middle-reg}
\end{figure}

\subsection{Dataset}\label{subsec:body}

The used dataset comprises audio 
recordings and ratings of auditions from the Florida 
Bandmaster’s Association (FBA) from 2013 to 2018. This 
dataset contains raw audio recordings from three 
different levels of  all-state auditions, \textit{\textit{Middle School}}, \textit{Concert Band}, 
and \textit{\textit{Symphonic Band}}. Each student performs
a prepared \textit{lyrical exercise}, \textit{technical exercise}, \textit{scales}, and a \textit{sight reading exercise}. This dataset includes several monophonic and percussive instruments. A subset of these 
data was used in this study, using
the technical exercise from the alto saxophone, Bb clarinet, and flute 
recordings for the \textit{Middle School} and \textit{Symphonic Band} levels. 
\tabref{tab:datapoints} shows the number of recordings per instrument for both \textit{Middle School} and \textit{Symphonic Band}. The average duration of a \textit{Middle School} and a \textit{Symphonic Band} recording is approximately \unit[30]{s} and \unit[50]{s}, respectively.

Each singular recording represents the complete audition for one student and has assessment ratings by an expert judge for four assessment criteria defined by the FBA: \textit{musicality}, \textit{note accuracy}, \textit{rhythm accuracy}, and \textit{tone quality}. For consistency, we normalized each rating to the range [0, 1] by dividing the maximum rating, with 0 representing the worst possible score, and 1 representing the best possible score. Furthermore, the tone quality rating was ignored for this study as the audition is represented as pitch contours at our network input, a representation that does not carry sufficient information for modeling this criterion. 

\subsubsection{Pre-processing}
Pitch contour representations were computed from raw audio recordings. Data augmentation via random chunking was used while training due to its ability to improve model performance\cite{pati_assessment_2018}. Each pitch contour is chunked into sections of length 1000 (about \unit[6]{s}) by randomly selecting the start position. This approach has been shown to improve model performance \cite{pati_assessment_2018}. We assume the chunked segment would receive the same assessment rating as the entire audition recording.


\subsection{Training procedures}
Pairs for the contrastive loss were randomly sampled via generated random sequences each batch. Each datapoint within the pair was fed into a separate encoder channel of the model.  
Each model was trained using a stochastic gradient descent optimizer with a weight decay of 1e-5 and momentum of 0.9. Early stopping was applied in each training sequence to stop if the validation loss had not decreased in 75 epochs. For training and evaluation, each dataset was split into training, testing, and validation sets in an 8:1:1 ratio. To measure the variance of the models, each model was trained five times using random seeds, as represented by the box plots. Within the two step training method following \cite{khosla_supervised_2020}, the encoder channels were trained for 150 epochs at a learning rate of 0.1, while the linear layers were trained for 300 epochs at a learning rate of 0.005. The Joint Loss Network was trained for 300 epochs at a learning rate of 0.005.

\begin{figure}
\centering
\includegraphics[width=\columnwidth]{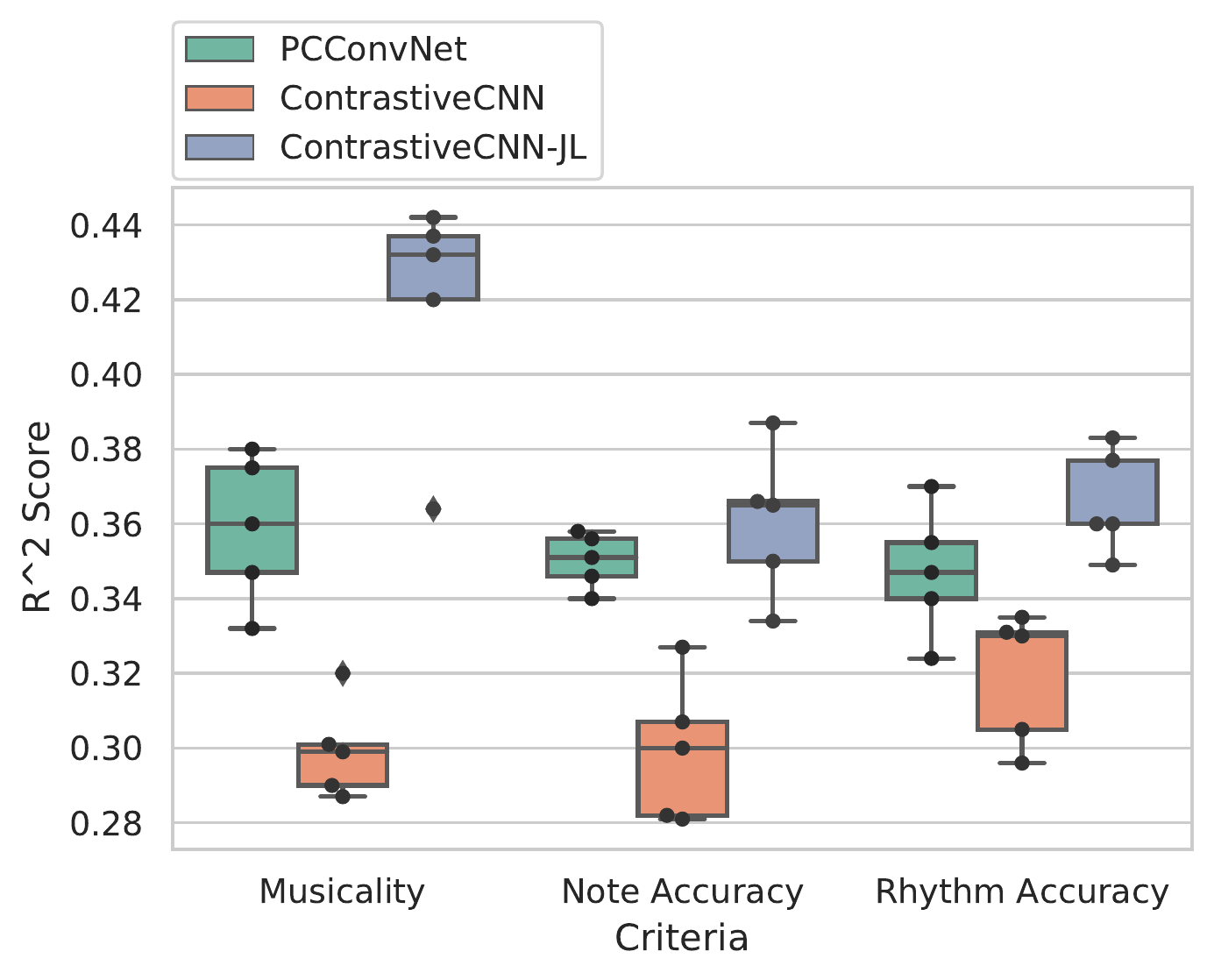}
\caption{Regression results over the \textit{Symphonic Band} set.}
\label{fig:symp-reg}  
\end{figure}

\subsection{Evaluation}
We investigate the performance of the baseline and the contrastive-regularized networks amongst three different rating criteria, \textit{musicality}, \textit{note accuracy}, and \textit{rhythm accuracy}. We predict the ratings for these criteria over both the \textit{Middle School} and \textit{Symphonic Band} dataset to determine performance over different levels of musical complexity, which can provide insights over the performance of MPA systems as musical complexity increases. The \textit{Concert Band} dataset was omitted for consistency, as it was not evaluated in previous MPA studies that used this dataset \cite{pati_assessment_2018, huang_score-informed_2020}. Each model (PCConvNet, ContrastiveCNN, ContrastiveCNN-JL) was trained separately for each assessment criterion on both datasets. The unaltered PCConvNet \cite{pati_assessment_2018} served as the baseline model. 
    \begin{figure*}
      \centering
      \begin{tabular}{@{}c@{}}
        \includegraphics[width=0.33\linewidth]{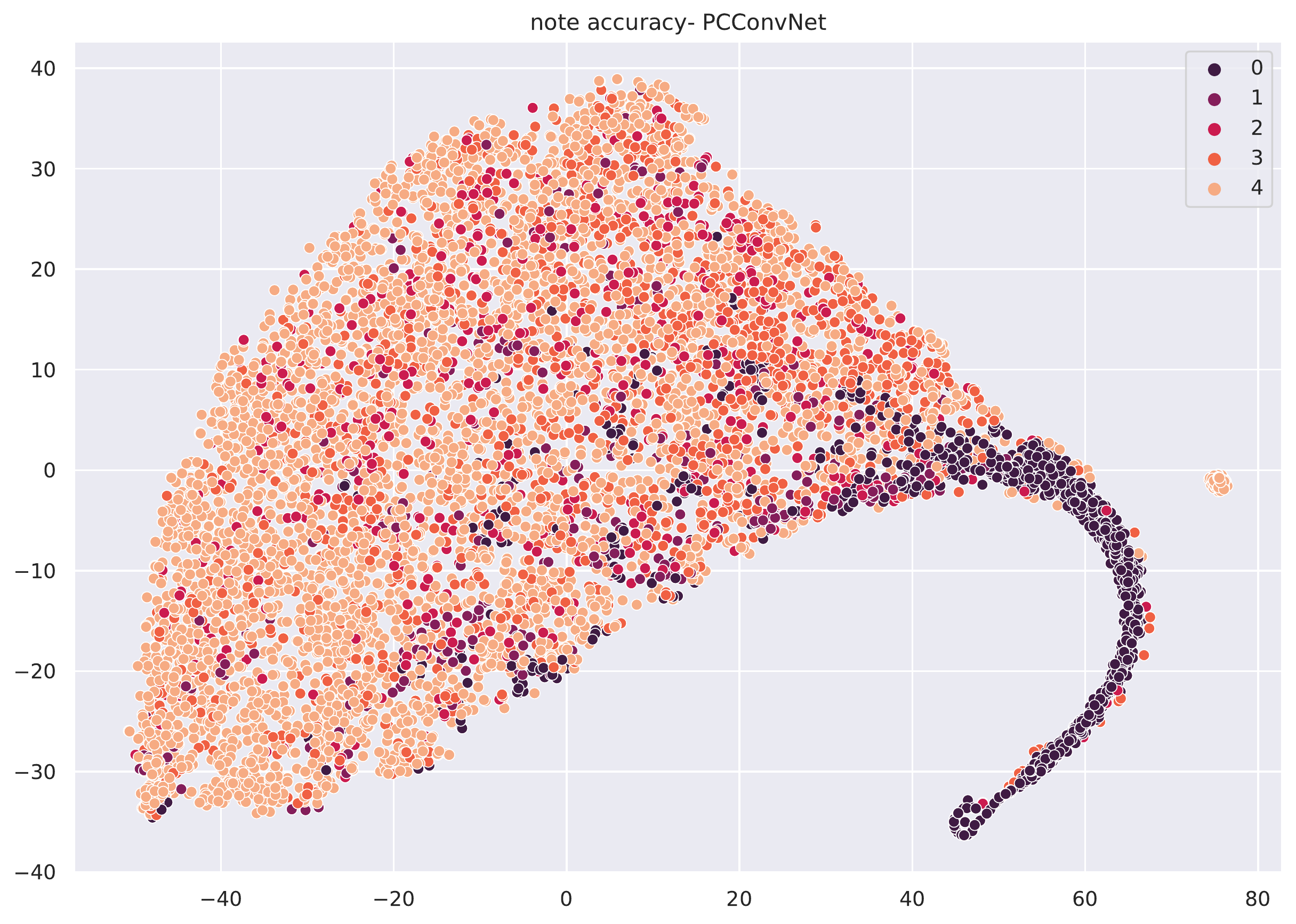} \\[\abovecaptionskip]
        \small (a) PCConvNet
      \end{tabular}
      \begin{tabular}{@{}c@{}}
        \includegraphics[width=0.33\textwidth]{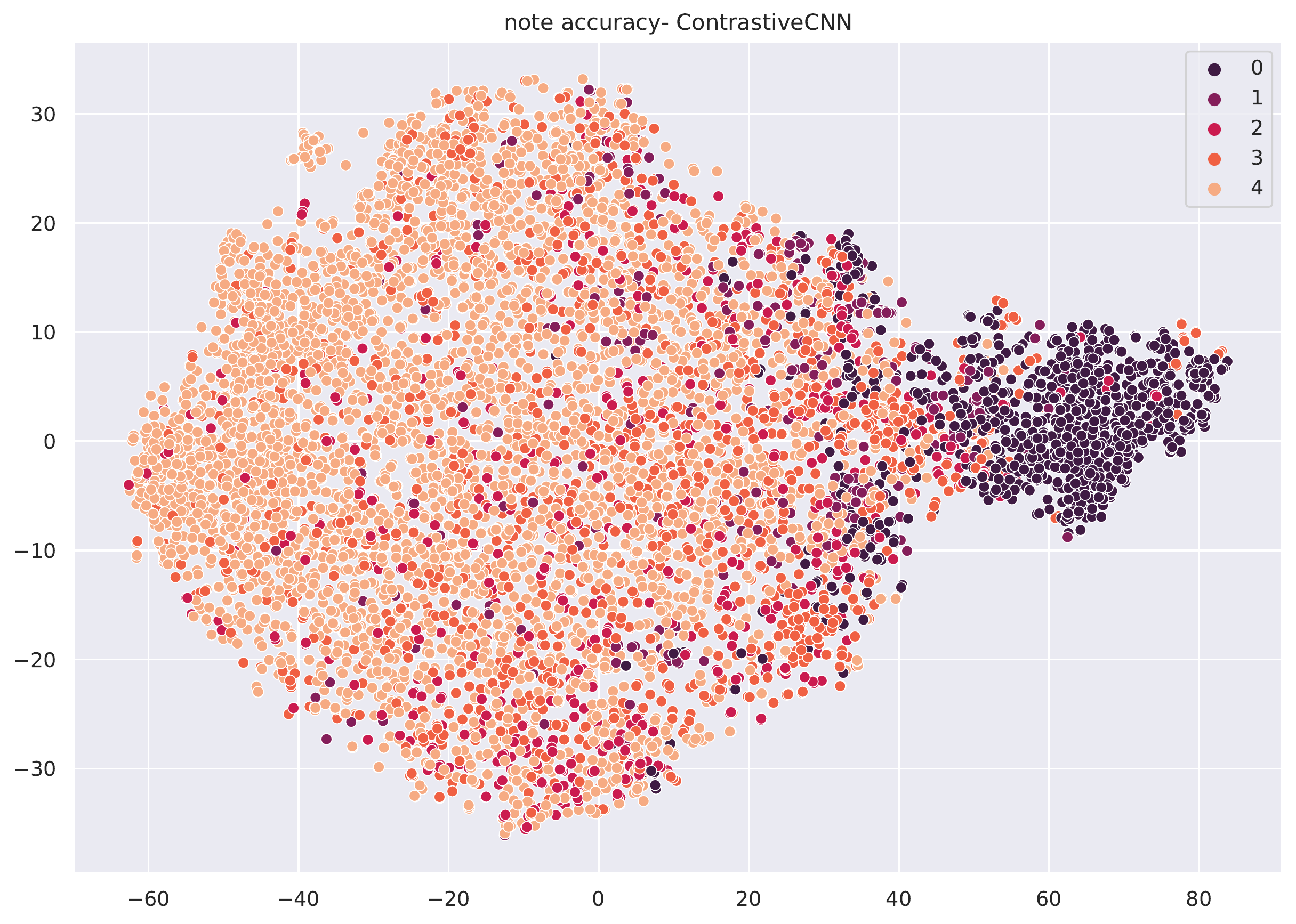} \\[\abovecaptionskip]
        \small (b) ContrastiveCNN
      \end{tabular}
      \begin{tabular}{@{}c@{}}
        \includegraphics[width=0.33\textwidth]{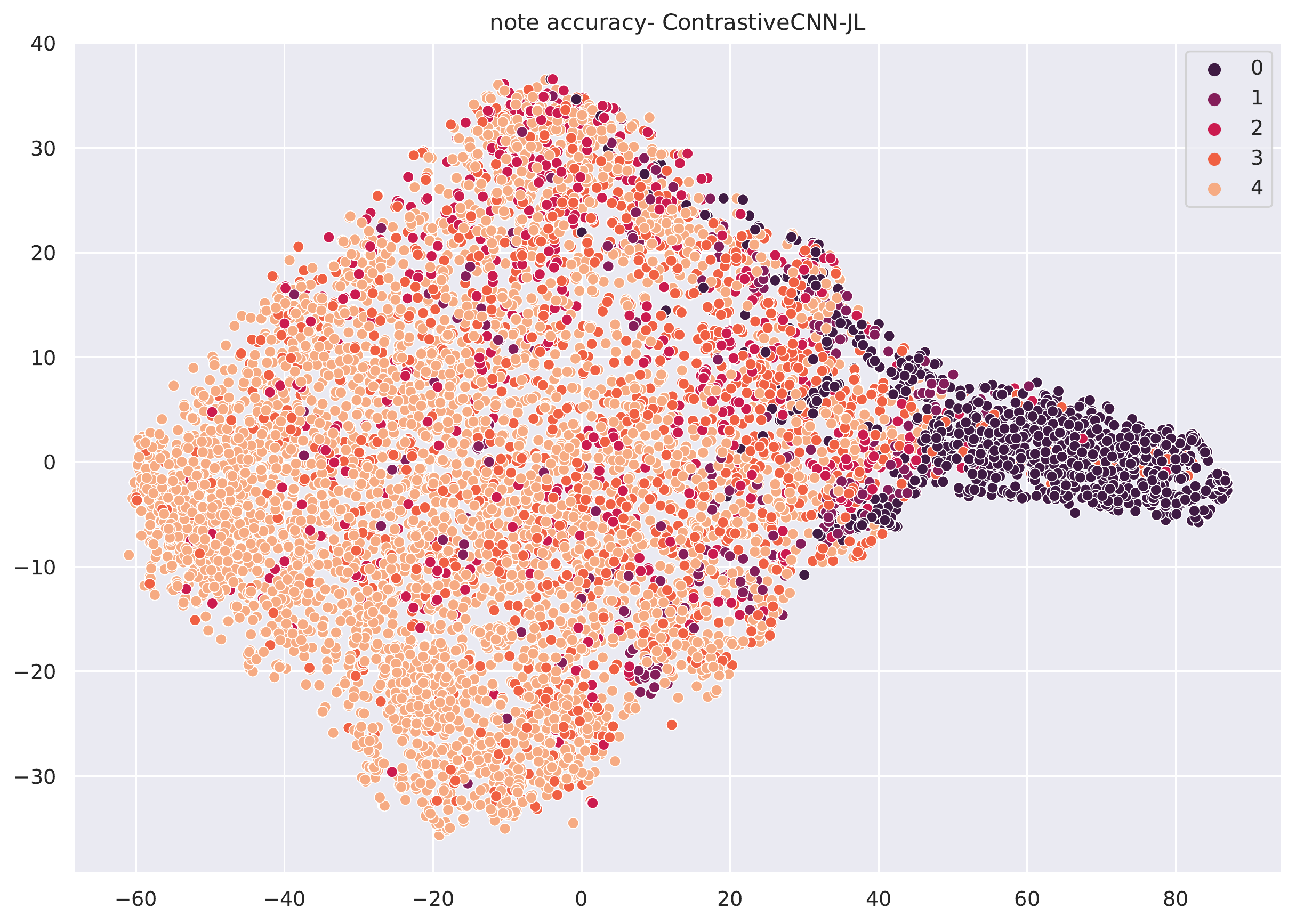} \\[\abovecaptionskip]
        \small (c) ContrastiveCNN-JL
      \end{tabular}
      \caption{T-SNE visualization of the latent space of the three presented models.} 
    \label{fig:latent}
    \vspace{1mm}
    \end{figure*}

\subsubsection{Regression analysis}
The coefficient of determination $(R^2 score)$ over the output scores serves as the evaluation metric:
\begin{equation}
    R^2 =1 -  \frac{\sum{_i}(y_i - \hat{y_i})^2}{\sum{_i}(y_i - \Bar{y})^2} ,
\label{eq:rs}
\end{equation}
%
where $y_i$ is the ground truth rating for a given datapoint, $\hat{y_i}$ is the predicted rating, and $\bar{y}$ is the mean ground truth rating over the entire set. 

\subsubsection{Latent space analysis}
Using each trained model, the latent vectors of the testing set were obtained by only passing the input through each model's encoder channels. A visualization was produced
by applying T-SNE dimensionality reduction \cite{van_der_maaten_visualizing_2008} to the latent vectors and plotting the output. Optimal parameters were found via a parameter search.

For a quantitative evaluation of the clustering quality, the latent space is evaluated by its  Davies-Bouldin index \cite{davies_cluster_1979}. 
The Davies-Bouldin index describes the average similarity of each cluster to its most similar cluster, which is defined as the ratio of within-cluster distances to between-cluster distances. The minimum index is zero and smaller values indicate better clustering \cite{davies_cluster_1979}.

\section{Results and discussion}
\label{sec:result}
\subsection{Regression results}
Figure~\ref{fig:middle-reg} and \figref{fig:symp-reg} detail the results of the models on the \textit{Middle School} and \textit{Symphonic Band} datasets, respectively. Each box plot contains the five runs with random seeds 0-4 per each criteria and model. We can make the following observations:
\begin{compactenum}[(i)]
    \item   All models perform better on the \textit{Middle School} set than the \textit{Symphonic Band} set (higher $R^2$ score). One possible explanation for this is that the \textit{Symphonic Band} auditions tend to be longer and more complex with higher skilled players, potentially increasing the difficulty of extracting meaningful features representing the quality of the performance. 
    \item   All contrastive-based models outperform the baseline on the \textit{Middle School} set; however, only the ContrastiveCNN-JL meets and outperforms the baseline on \textit{Symphonic Band}. This implies that the contrastive learning is more beneficial at a lower complexity of performance, but possibly has difficulty with data of higher complexity. One possible explanation could be that with a higher level of performance, and thus a higher complexity of information within each latent vector, the contrastive loss is unable to properly semantically relate the distances to the quality of the performance.
    \item   The ContrastiveCNN-JL outperforms both the baseline and the ContrastiveCNN in every trial. This implies that the information gained by combining the traditional loss term with the contrastive loss helps learning a more meaningful latent space representation.
\end{compactenum}


\subsection{Latent space analysis}
\subsubsection{T-SNE plots}
T-SNE  visualizations of the latent space are presented for the baseline PCConvNet, ContrastiveCNN, and ContrastiveCNN-JL in \figref{fig:latent}. As a example, we only present results for \textit{Note Accuracy} on the \textit{Middle School} dataset. The effect of the contrastive loss can be easily noticed, although the embedding spaces are not ordered perfectly in either case. The two models based on contrastive loss display a more defined distinction between low classes (0, 1) and higher classes (3, 4). Small same-class clusters can also be identified. 

\subsubsection{Class Distance Surface plots}
\begin{figure*}
      \centering
      \begin{tabular}{@{}c@{}}
        \includegraphics[width=0.33\linewidth]{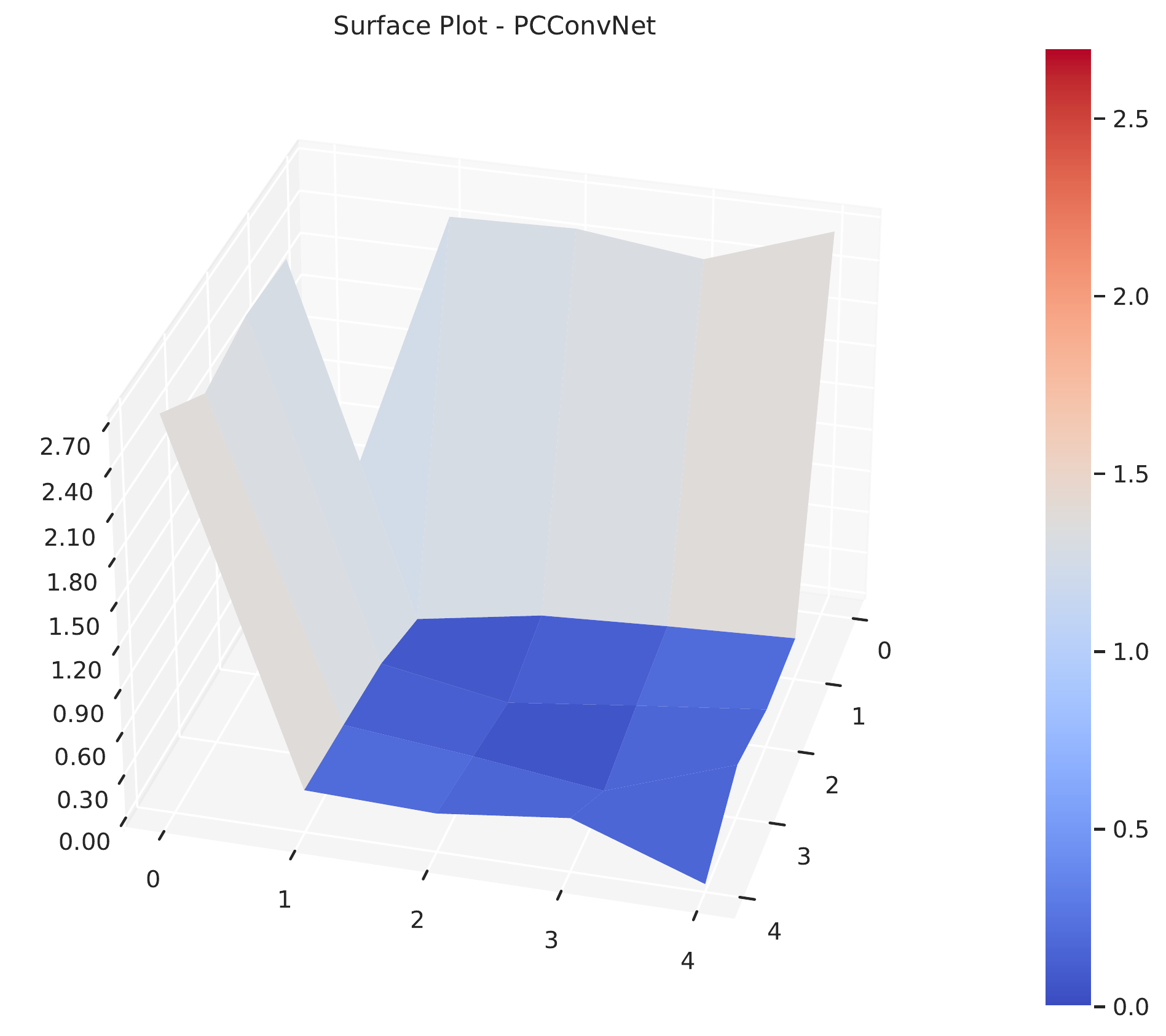} \\[\abovecaptionskip]
        \small (a) PCConvNet
      \end{tabular}
      \begin{tabular}{@{}c@{}}
        \includegraphics[width=0.33\textwidth]{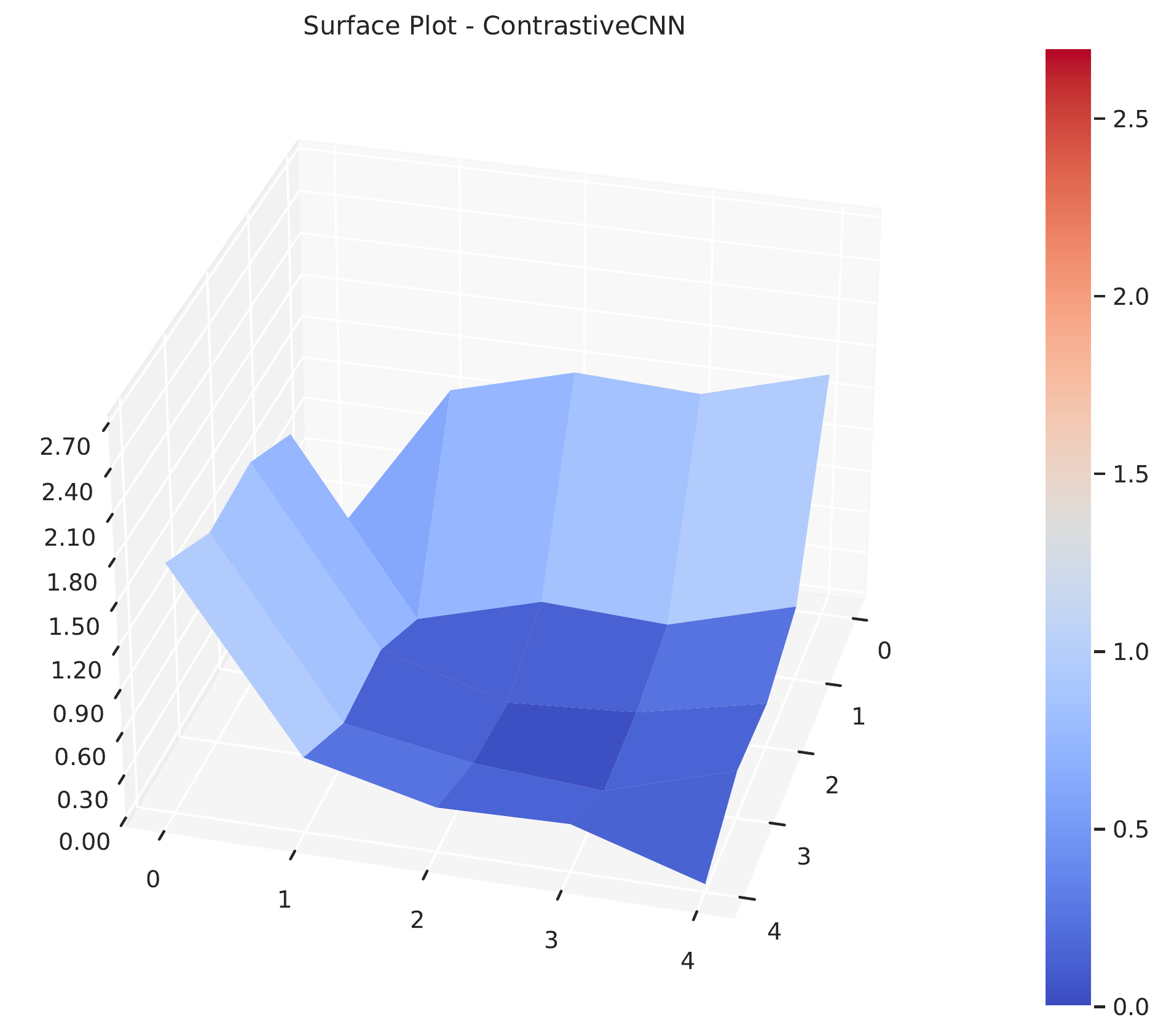} \\[\abovecaptionskip]
        \small (b) ContrastiveCNN
      \end{tabular}
      \begin{tabular}{@{}c@{}}
        \includegraphics[width=0.33\textwidth]{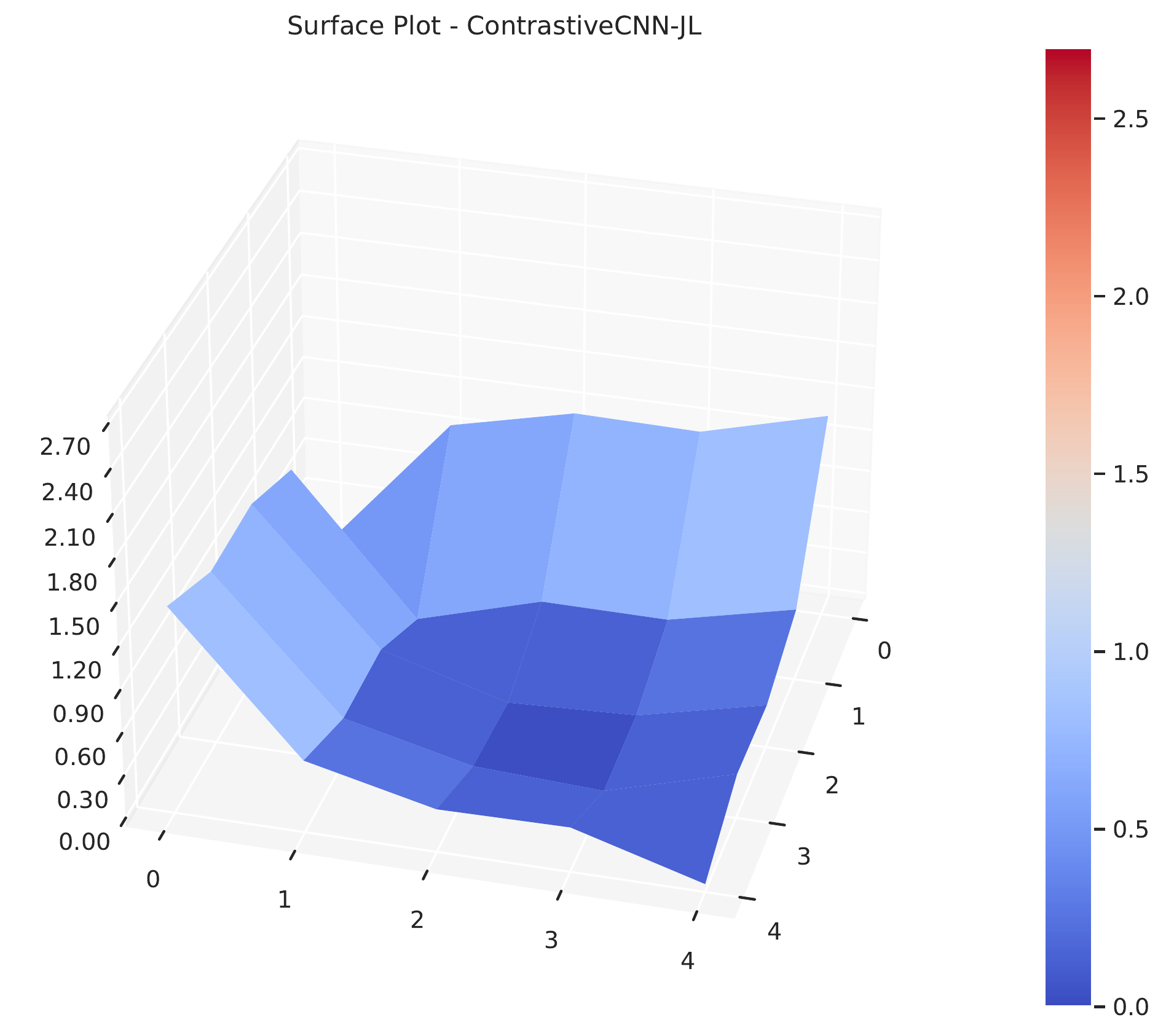} \\[\abovecaptionskip]
        \small (c) ContrastiveCNN-JL
      \end{tabular}
      \caption{Class distances in the learned latent space of the three models.} 
    \label{fig:surf}
    \end{figure*}

\figref{fig:surf} shows the distances between the centroids of each class cluster within the latent space of the models trained on the \textit{Middle School} dataset for \textit{Note Accuracy}. While the contrastive-based models appear to have trouble properly ordering the middle range of ratings between classes 2 and 3, the distances appear to scale more smoothly than the distances within the baseline PCConvNet, indicating better ordering within the latent space.

\subsubsection{Davies-Bouldin Index}
\begin{figure}
\centering
\includegraphics[width=75mm]{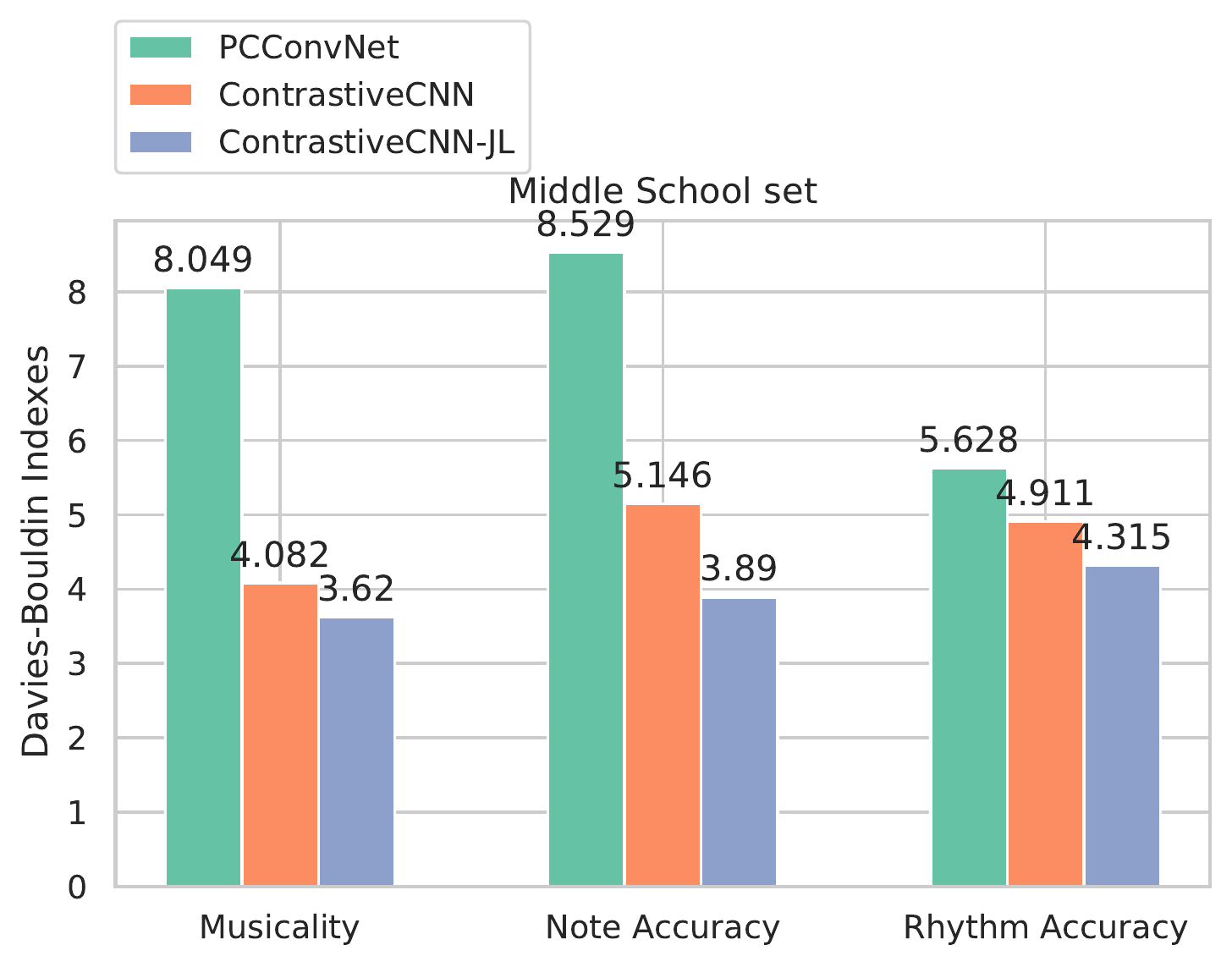}
\caption{Davies-Bouldin indices of each model over the \textit{Middle School} dataset. Smaller values indicate better clustering.}
\label{fig:mid-db}  
\end{figure}

\figref{fig:mid-db} shows the Davies-Bouldin indices of each model on the \textit{Middle School} regression set. Each contrastive-based model has a considerably lower index than the baseline, which indicates that the latent space clustering is improved. Within this set, the lower the Davies-Bouldin index, the better the regression performance, implying that better clustered latent spaces do correlate with better regression.

\section{Conclusion}
This paper presented an approach to representation learning to improve the accuracy of a system for music performance assessment. We introduced a weighted contrastive loss suitable for regression tasks and showed how this latent space regularization improves results on a large real-world dataset for music performance assessment.


In future work, 
we plan to incorporate score information into the models, as this has been shown to improve performance \cite{huang_score-informed_2020}. More analysis should be done within contrastive learning methods to assess the effect of margin size, and number of classes on the performance of the model and the goodness of its clustering. 
Another approach to ensure that the learned representations contain relevant information is multi-task learning. It is worth investigating what related tasks might help increase the performance of music performance assessment. 
Moreover, supervised latent space regularization methods such as AR-VAE \cite{pati_attribute-based_2020} and I-VAE \cite{khemakhem_variational_2020} might be incorporated to force specific dimensions to specific performance characteristics.

\section{Acknowledgments}
We would like to thank the Florida Bandmasters Association for providing the dataset used in this study.

We also gratefully acknowledge NVIDIA Corporation (Santa
Clara, CA, United States) who supported this research via the NVIDIA GPU Grant
program.

\bibliography{contrastivempa}

\end{document}